\documentclass[prl,longbibliography,twocolumn]{revtex4-1}
\usepackage{geometry}
\usepackage{amsmath}
\usepackage{amssymb}
\usepackage{wrapfig}
\usepackage{subcaption}
\usepackage{graphicx,picture,calc}
\usepackage{hyperref}
\usepackage{braket}
\usepackage{bm}
\usepackage[normalem]{ulem}
\usepackage{xcolor}
\usepackage{epstopdf}

\usepackage{natbib}

\newcommand{\mb}{\mathfrak{b}}

\newcommand{\be}{\begin{equation}}
\newcommand{\ee}{\end{equation}}
\newcommand{\bk}{{{\bf{k}}}}

\newcommand{\bQ}{{{\bf{Q}}}}
\newcommand{\br}{{{\bf{r}}}}

\newcommand{\bq}{{\bf{q}}}

\newcommand{\bea}{\begin{eqnarray}}
\newcommand{\eea}{\end{eqnarray}}
\newcommand{\beal}{\begin{align}}
\newcommand{\eeal}{\end{align}}
\renewcommand{\i}{{\mathrm{i}}}
\newcommand{\ra}{\rangle}
\newcommand{\la}{\langle}

\newcommand{\upa}{\uparrow}
\newcommand{\dna}{\downarrow}
\newcommand{\bS}{{\bf S}}

\newcommand{\dg}{{\dagger}}
\newcommand{\pdg}{{\phantom\dagger}}

\def\l{\ell}

\begin{document}
\title{Magnetic and Nematic Orders of the Two-Dimensional Electron Gas at Oxide (111) Surfaces and Interfaces}
\author{Nazim Boudjada}
\affiliation{Department of Physics, University of Toronto, Toronto,
  Ontario M5S 1A7, Canada}
\author{Gideon Wachtel} 
\affiliation{Department of Physics, University of Toronto, Toronto,
  Ontario M5S 1A7, Canada}
\author{Arun Paramekanti} 
\affiliation{Department of Physics, University of Toronto, Toronto,
  Ontario M5S 1A7, Canada}
\begin{abstract}
  Recent experiments have explored two-dimensional electron gases (2DEGs) at
  oxide (111) surfaces and interfaces, finding evidence for
  hexagonal symmetry breaking in SrTiO$_3$ at low temperature. 
  We discuss many-body
  instabilities of such (111) 2DEGs, incorporating multiorbital
  interactions in the $t_{2g}$ manifold which can induce
  diverse magnetic and orbital orders. Such broken symmetries may
  partly account for the observed nematicity, cooperating or competing with
phonon mechanisms. We present an effective field theory for the
  interplay of magnetism and nematic charge order, and discuss
  implications of the nematicity for transport and
  superconductivity in (111) 2DEGs.
\end{abstract}

\maketitle

{\it Introduction.---} Transition metal oxide heterostructures and
interfaces can realize exotic low-dimensional electronic phases and
allow for engineering oxide-based devices \cite{Review2014}. Extensive
research
\cite{ohtomo2004high,thiel2006tunable,nakagawa2006some,Millis2006,
  reyren2007superconducting,caviglia2008electric,caviglia2010tunable,Ariando2011,
  bert2011direct,li2011coexistence, macdonald_001,mehta2012evidence,held_001,
  fischer_spinorbit_2013, kim2013origin, banerjee2013ferromagnetic, chen2013,
  Park2013,ruhman_competition_2014,caviglia_001,Taraphder2016,
  tolsma_orbital_2016,atkinson2017influence} has focused on the two-dimensional electron gas
(2DEG) at the (001) LaAlO$_3$-SrTiO$_3$ (LAO-STO) interface induced by
a combination of the polar catastrophe and oxygen vacancies.  This
2DEG shows evidence of correlated magnetism in torque
magnetometry and scanning SQUID measurements
\cite{bert2011direct,li2011coexistence}. In addition, it exhibits superconductivity
(SC) \cite{reyren2007superconducting,PhysRevLett.107.056802} which may be tied to that
of doped bulk STO, though the interface might harbor
modulated SC pairing \cite{Michaeli2012} or Majorana modes
\cite{Fidkowski2013}.

Recently, various groups have started to probe 2DEGs at oxide (111)
surfaces and interfaces, for instance induced by photon
\cite{rodel_orientational_2014}
or ion \cite{mckeown_walker_control_2014, miao_anisotropic_2016} irradiation at the (111) STO
surface, as well as that at the (111) LAO-STO interface
\cite{rout_six-fold_2017,PhysRevB.95.035127,ADMI:ADMI201600830,monteiro_two-dimensional_2017, davis_superconductivity_2017}.
Part of this interest stems from proposals for realizing topologically nontrivial phases along this growth direction \cite{Xiao_NComm2011,Ruegg_PRB2011,Cook_PRL2014,Okamoto_PRB2014,Fiete_SciRep2015,Baidya_PRB2016,Held_2016,Kee_NPJQM2017}.
The [111] growth direction is polar for STO due to alternating
Ti$^{4+}$ and (SrO$_3)^{4-}$ layers, and the internal electric
fields could lead to stronger confinement \cite{baidya2015} of the
(111) 2DEG, potentially enhancing correlation effects relative to (001) 2DEGs.
Angle resolved photoemission spectroscopy (ARPES) on
the (111) STO surface reveals a Fermi surface (FS) composed of all three
$t_{2g}$ orbitals, which appears to preserve the expected hexagonal
symmetry \cite{rodel_orientational_2014, mckeown_walker_control_2014}.
However, very recent experiments have discovered, via measurements of
magnetotransport \cite{miao_anisotropic_2016,rout_six-fold_2017,PhysRevB.95.035127,ADMI:ADMI201600830} and the
resistive transition into the SC state
\cite{davis_superconductivity_2017}, that this (111) 2DEG exhibits an
anisotropy which sets in at low temperatures, spontaneously breaking
the hexagonal symmetry. While a weak resistive anisotropy may arise from
the ${\sim 100}\;\rm K$ pseudo-cubic to pseudo-tetragonal transition
of bulk STO \cite{aharony_trigonal--tetragonal_1977}, the onset
temperature seen in these experiments is much lower, ${T_{\rm ani}\sim
  4\text{-}30\;\rm K}$ depending on the sample and the electron
density. For bulk STO, it is known that the transition into the
pseudo-tetragonal phase is sensitive to stress along the [111]
direction \cite{muller_indication_1991}, and proceeds via an
intermediate trigonal phase; it remains to be tested if the lower
symmetry at the (111) surface or interface leads to a low
temperature surface phonon instability.

In light of these developments, it is in any case also important to
consider the impact of electron-electron interactions on (111) 2DEGs,
in order to (i) study possible interaction induced many-body
instabilities, and (ii) ask if there are electronic mechanisms for the
observed anisotropies of the (111) 2DEG which may cooperate or compete
with phonon instabilities. Such an interplay has been actively investigated in the iron
pnictide superconductors (see Ref.~\onlinecite{FernandesReviewPnictide2014}
for a review).

\begin{figure}[t]
  \centering
  \includegraphics[width=\linewidth]{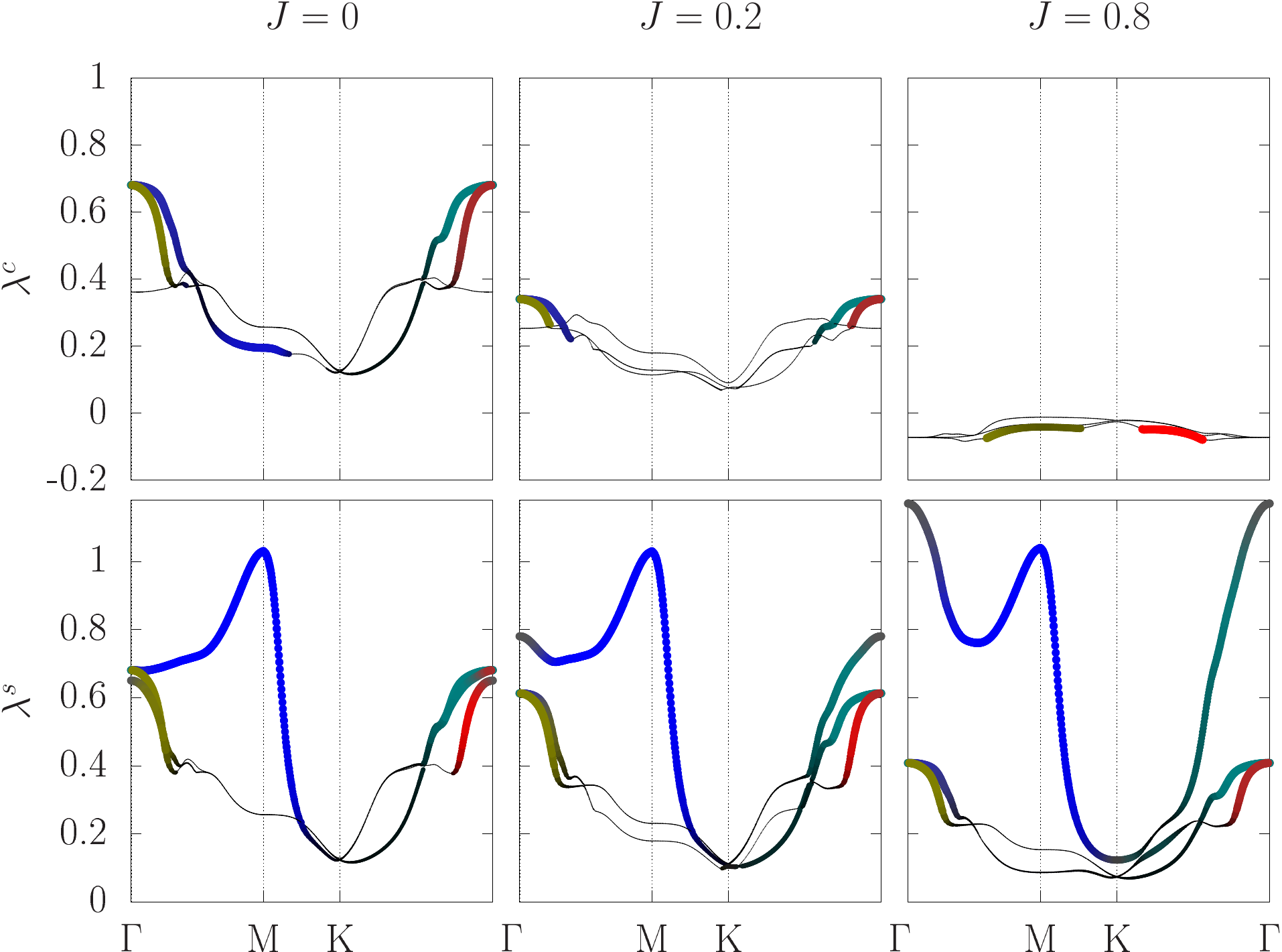}
  \caption{Dominant three eigenvalues $\lambda^{(c,s)}$ ($c=$ charge, $s=$ spin) of the
    matrix product
    $\underline{U}^{(c,s)}\underline{\chi}_0(\bq,\Omega=0)$, plotted
    along high symmetry lines of the BZ, 
    with $t = 1$, $U = 2$,
    $t_\perp=-t'=0.04$, and $T=0.02$, for fixed density $\bar{n}=0.3$, and
    varying Hund's coupling $J/t$. Line color (red,green,blue)
    indicates relative weight of orbitals (respectively $xy,xz,yz$) in the orbital-diagonal part of the 
    eigenvectors $|f^{(c,s)}_{\l\l}|^2$; thickness indicates
    total weight, $\sum_\l|f^{(c,s)}_{\l\l}|^2$. The dominant instability
    is in the spin channel, being antiferromagnetic (near M) for small $J$
    and ferromagnetic (at $\Gamma$) for large $J$. 
    In the charge channel, the leading instability at small $J$ is a two-fold degenerate 
    mode at $\Gamma$ corresponding to ferro-orbital order.}
  \label{fig:rpa}
\end{figure}

\begin{figure*}[t]
	\centering
	\includegraphics[width= 0.24\linewidth]{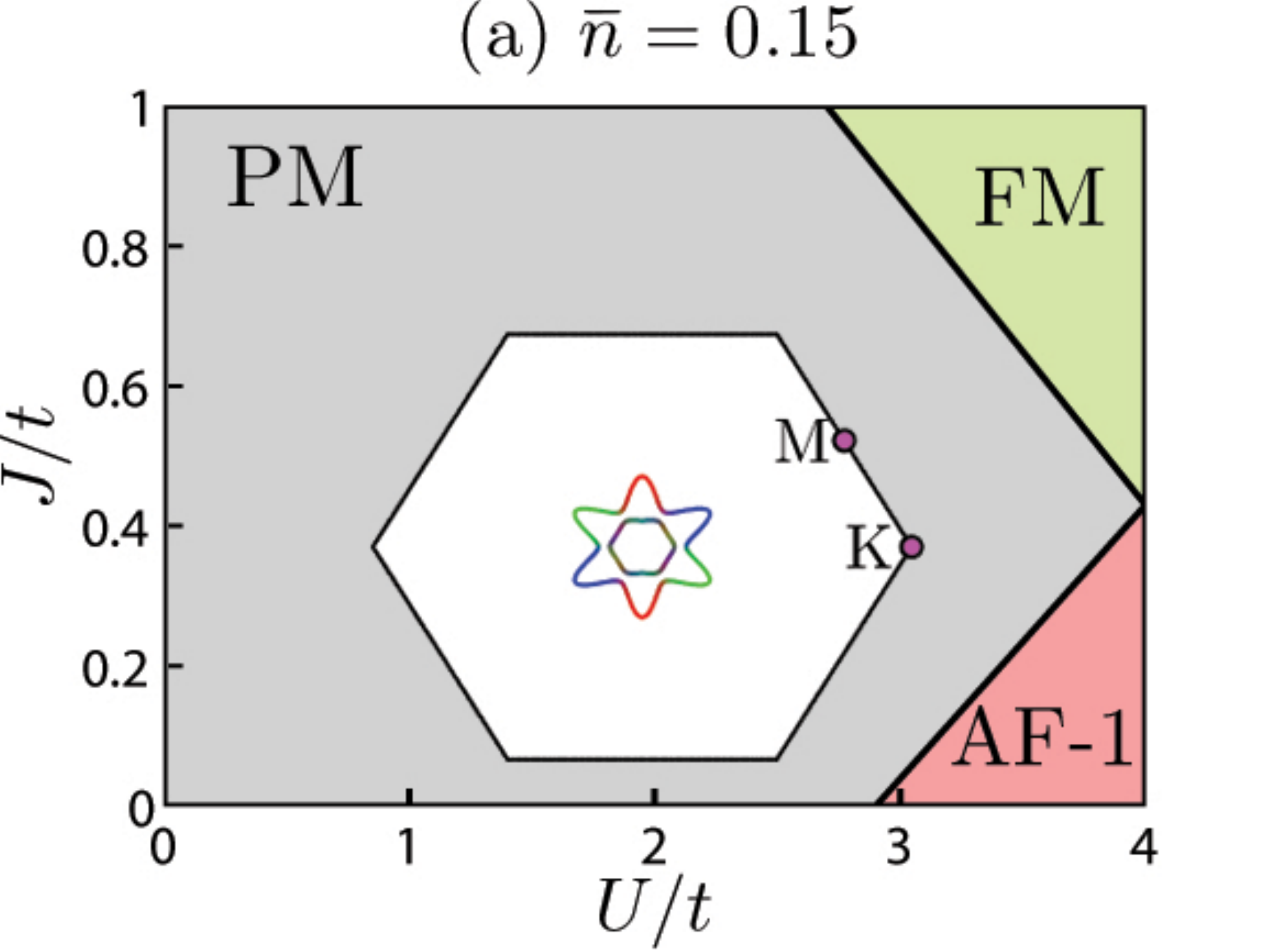}
	\includegraphics[width= 0.24\linewidth]{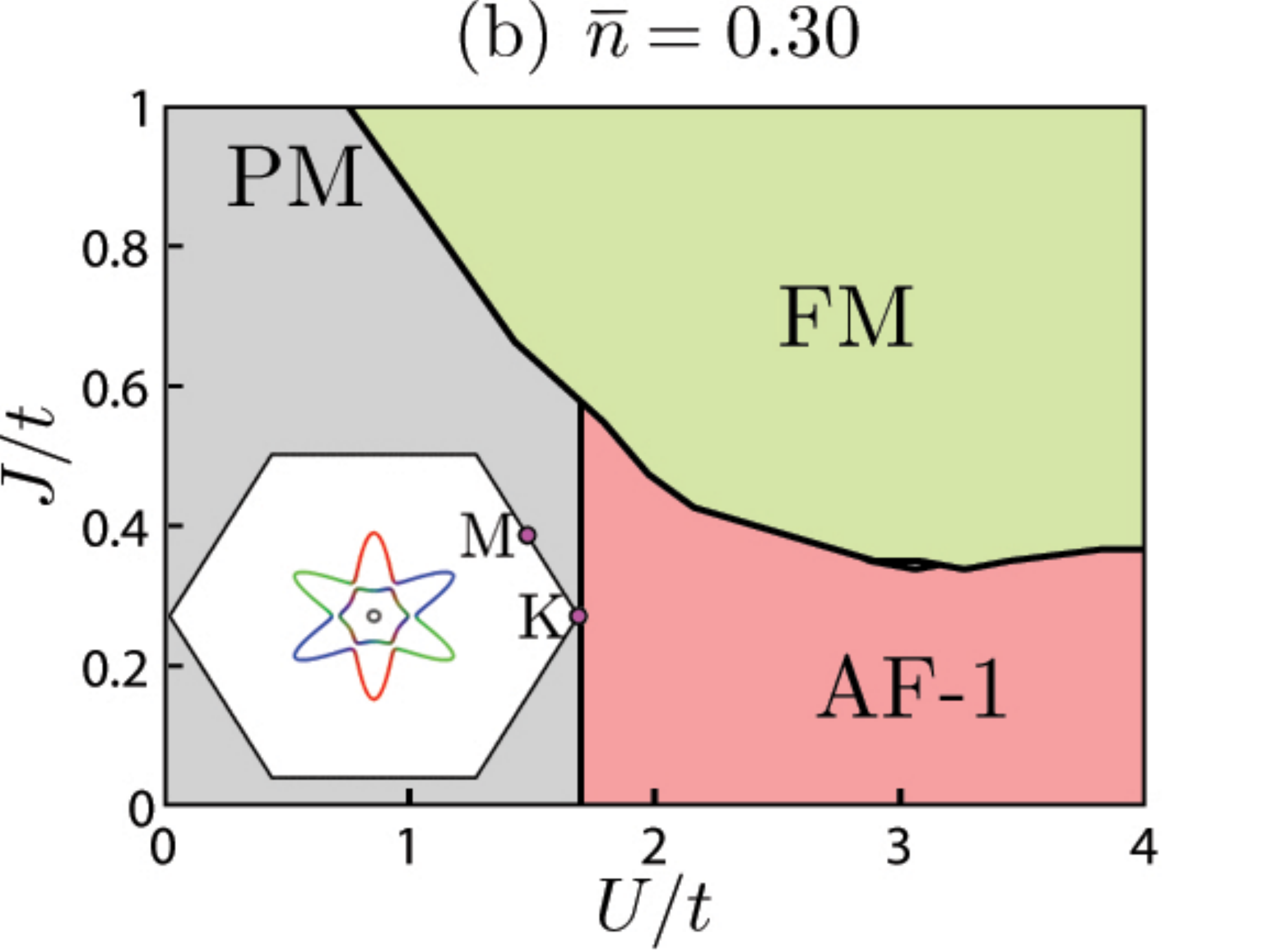}
	\includegraphics[width= 0.24\linewidth]{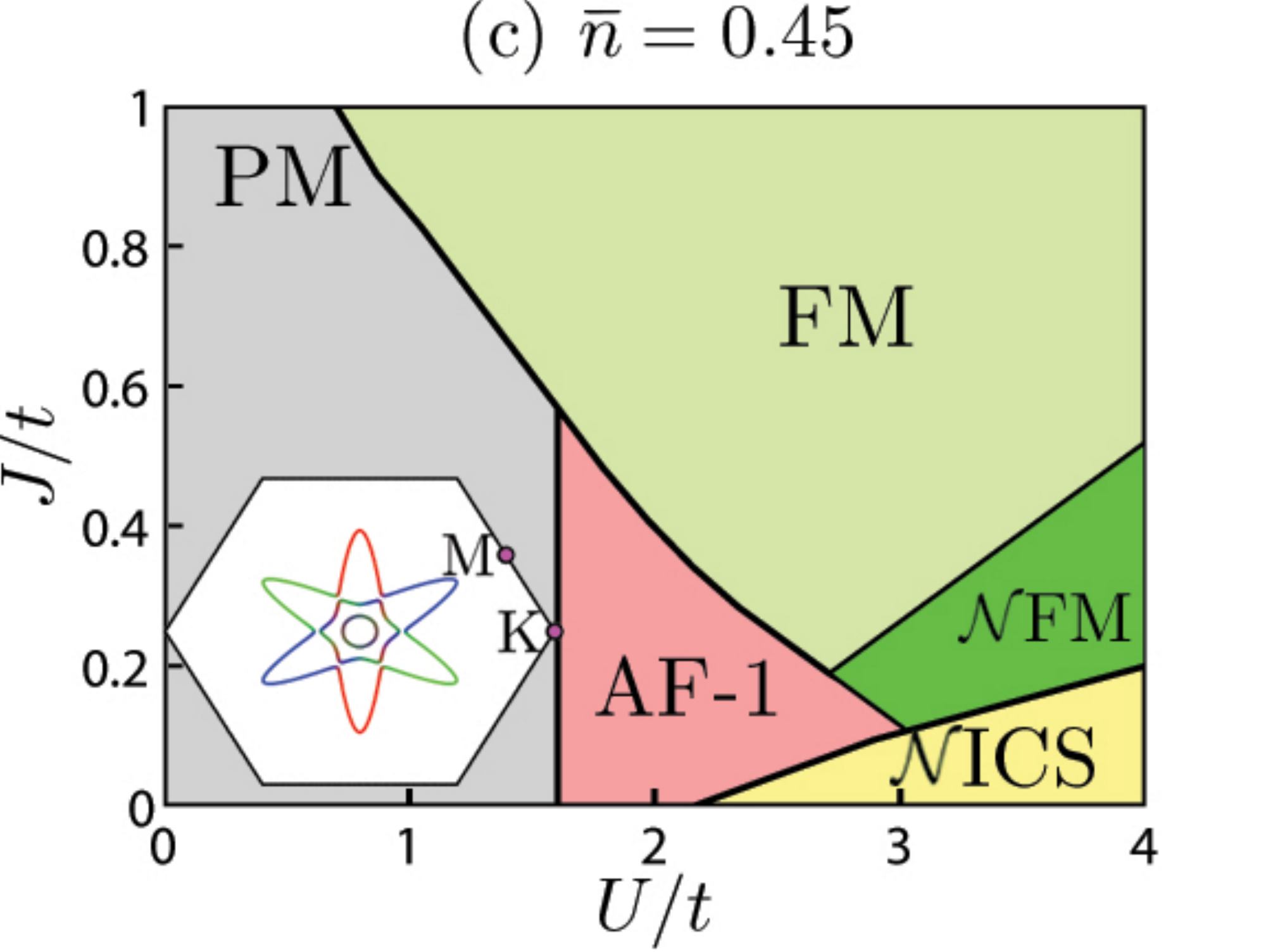}
	\includegraphics[width= 0.24\linewidth]{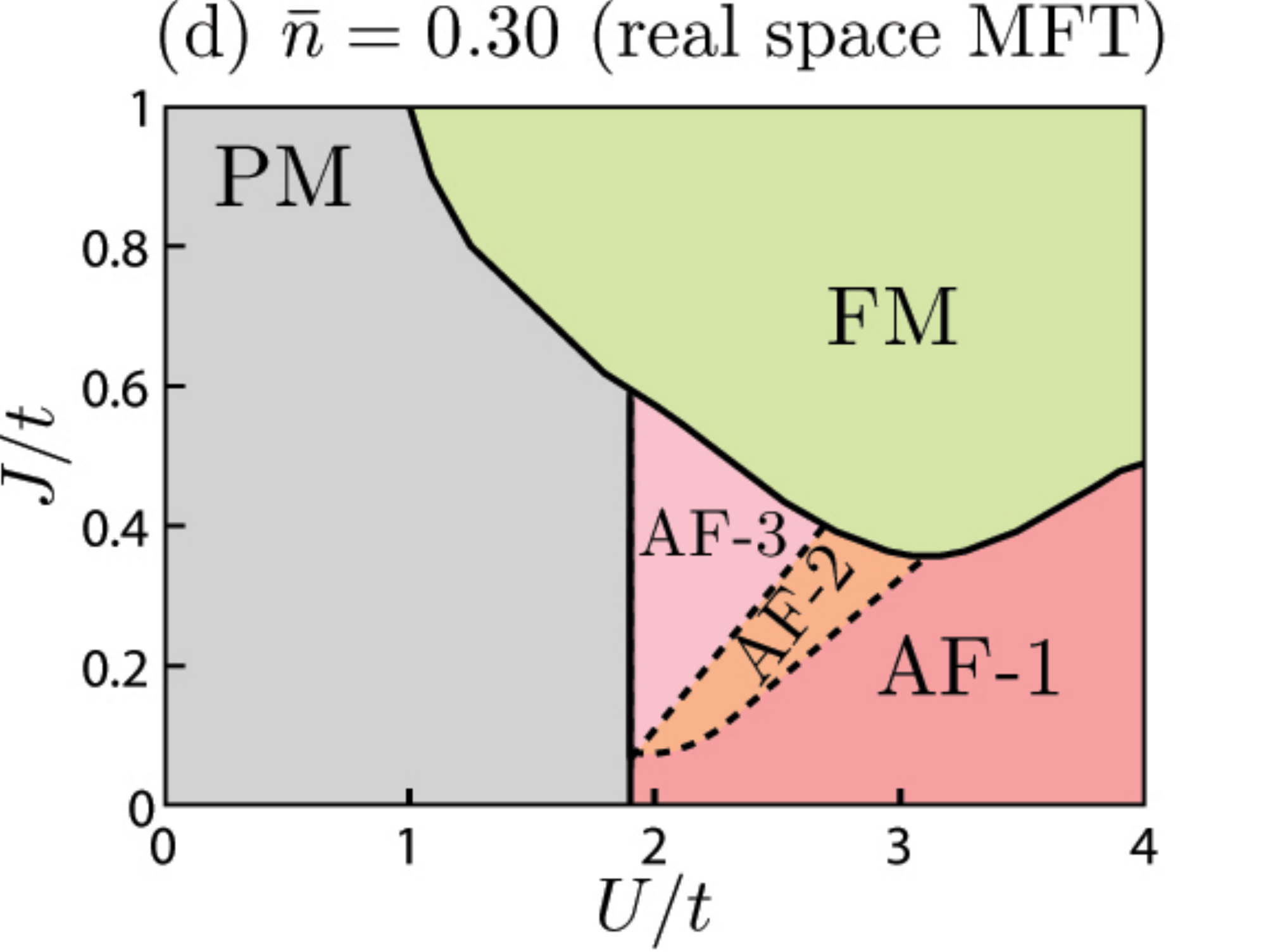}
	
	\caption{Zero temperature phase diagram of the (111) 2DEG as a function of the Hubbard repulsion $U/t$ and
			Hund's coupling $J/t$ for densities (a) $\bar{n}=0.15$, (b) $\bar{n}=0.30$, (c) $\bar{n}=0.45$ within a single-$\bQ$ spiral
			mean field theory (MFT).  The different
			metallic phases are paramagnetic phase (PM), stripe antiferromagnet (AF-1), incommensurate spiral (ICS),
			and a ferromagnetic phase (FM) at large Hund's coupling. The PM phases depict
			the noninteracting Fermi surfaces. (d) Real space
			MFT (at $\bar{n}=0.3$, $T=0.02t$) showing that the commensurate AF might support regimes of multi-$\bQ$ orders (AF-2, AF-3). 
			The $\mathcal{N}$FM, ICS, AF-1, and AF-2 phases coexist with ferro-orbital order which will lead to
			transport anisotropies.}
	\label{fig:phasediagram}
\end{figure*}

Motivated by these questions, we examine
a model for $t_{2g}$ electronic states of the (111) STO surface
2DEG, which is consistent with the ARPES measurements, and study its
instabilities driven by multiorbital interactions. Our main findings,
summarized in Figs.~\ref{fig:rpa} and \ref{fig:phasediagram}, based on a combination of
random phase approximation (RPA) calculations supplemented by mean
field theory, is that there is a range of densities over which this
2DEG is unstable to ferromagnetic (FM) or antiferromagnetic (AF)
order, accompanied by
ferro-orbital order. Even if thermal
fluctuations melt such magnetic orders in 2D, the
orbital order and the fluctuating magnetism are expected to
survive to higher temperatures, leading to a nematic fluid
\cite{oganesyan2001quantum,khavkine2004formation,
  kee2005itinerant,fradkin2007electron,fang2008theory,fradkin2010nematic}
which breaks hexagonal symmetry.
We present a
Landau theory of this nematic, and discuss implications
for transport measurements and superconductivity. Such nematicity
induced by orbital or spin order has been previously considered for
the (001) 2DEG \cite{PhysRevB.91.241302,Fischer2013,XLi2014,tolsma_orbital_2016}. 
Our
results should be broadly applicable to a wide class of oxide (111) 2DEGs.

{\it Model. ---}
We begin with a tight-binding model of Ti $t_{2g}$-orbitals on a 2D triangular lattice which captures the FS seen in ARPES 
 \cite{rodel_orientational_2014, mckeown_walker_control_2014}
for 
the (111) 2DEG at the STO surface:
\be
  \label{eq:H0}
  H_0 = \sum_{\bk\ell\ell'\sigma} c^\dg_{\ell\sigma}(\bk) h^\pdg_{\ell\ell'}(\bk) c^\pdg_{\ell'\sigma} (\bk)
\ee
with $\ell \equiv yz,zx,xy$, and
\bea
  \label{eq:h}
\underline{h}(\bk) = \begin{pmatrix}
    \epsilon^{c}_\bk + \eta^{ab}_\bk &  \gamma^a_\bk & \gamma^b_\bk \\
    \gamma^a_\bk & \epsilon^{b}_\bk + \eta^{ca}_\bk   & \gamma^c_\bk \\
\gamma^b_\bk & \gamma^c_\bk & \epsilon^a_\bk + \eta^{bc}_\bk
\end{pmatrix}, \eea 
where $\epsilon^{\alpha}_\bk \!=\! -2 t \cos k_\alpha$ and
$\eta^{\alpha\beta}_\bk \!=\! -2 t_\perp (\cos k_\alpha \!+\! \cos
k_\beta)$ determine the intraorbital dispersion which leads to elliptical FSs, while $\gamma^\alpha_\bk = -
2 t' \cos k_\alpha$ captures weak interorbital hopping. Here, we have
defined $k_\alpha = \bk\cdot \hat{\alpha}$ ($\alpha=a,b,c$), with
$\hat{a}=\hat{x}$, $\hat{b}=\hat{x}/2+\hat{y} \sqrt{3}/2$, and
$\hat{c}=-\hat{x}/2+\hat{y} \sqrt{3}/2$. We work in units where the
triangular lattice constant $d \approx 5.66\;\rm\AA$ is set to unity. 
The ARPES data  \cite{rodel_orientational_2014, mckeown_walker_control_2014} 
can be reasonably fit by choosing ${t = 320\;\text{meV}}$
and $t_\perp = 0.04 t$, and an electron density of $\bar{n} = 0.3$
electrons per site, corresponding to $10^{14}$ cm$^{-2}$; we therefore study a range of densities around this value. 
The interorbital terms appear to be small; for concreteness, we
set $t'=-0.04 t$. The resulting FSs are shown overlaid on the paramagnetic phases in
Fig.~\ref{fig:phasediagram}. The real 2DEG
wave functions will be spread over a few layers, so $H_0$ should only
be viewed as the simplest 2D tight-binding parameterization of the
ARPES data.  We omit spin-orbit coupling (SOC), but
comment on its effects later.  The local multiorbital interactions are \bea
  \label{eq:Hint}
  H_{\rm int} &=& U\sum_{i\ell}n^\pdg_{i\ell\uparrow}n^\pdg_{i\ell\downarrow}+\frac{1}{2} 
  V \sum_{i\ell\ne\ell'}n^\pdg_{i\ell}n^\pdg_{i\ell'} \nonumber \\
  &-&\!J\sum_{i\ell\ne\ell'}\!\bS^\pdg_{i\ell}\!\cdot\!\bS^\pdg_{i\ell'}
  + J' \sum_{i\ell\ne\ell'}
  c^\dagger_{i\ell\upa}c^\dagger_{i\ell\dna}c^\pdg_{i\ell'\dna}c^\pdg_{i\ell'\upa}.
  \eea 
where $i$ denotes the site and $\ell$ the orbital. Below, we fix $V\!=\!(U-5 J/2)$ and $J'\!=\!J$ 
as appropriate for $t_{2g}$ orbitals,
and explore broken symmetry states driven by varying the interactions $J/t,U/t$. These
interactions should be scaled down compared to atomic values by the
number of layers over which the 2DEG is spread.

{\it RPA analysis. ---}
To identify the leading weak-coupling instabilities we use an unbiased
multi-orbital RPA approach \cite{graser_near-degeneracy_2009}, with
the matrix response
${\underline{\chi}^{(c,s)}_{RPA}(\bq,\Omega)=\underline{\chi}_0(\bq,\Omega)
  (1-\underline{U}^{(c,s)}\underline{\chi}_0(\bq,\Omega))^{-1}}$, where
\begin{eqnarray}
  \label{eq:chi0}
  & & \left(\underline{\chi}_0(\bq,\Omega)\right)_{\l_1\l_2;\l_3\l_4} = 
  \frac{1}{N}\sum_{ij\sigma\sigma'}\int_0^\beta \mathrm{d}\tau
  \,\mathrm{e}^{\i\bq\cdot(\br_i-\br_j)-\i\Omega\tau} \nonumber \\ & & 
  \qquad\qquad\quad\times\left\langle
    c_{i\l_1\sigma}^\dagger(\tau) c_{i\l_2\sigma}(\tau)c_{j\l_3\sigma'}^\dagger(0) c_{j\l_4\sigma'}(0)
    \right\rangle.
\end{eqnarray}
is the bare response function (see Supplemental Material (SM)).
Here, $N$ is the number of sites, $c$ and $s$ respectively denote charge and spin responses. The
non-zero interaction vertex matrices are,
\begin{eqnarray}
  & &
  \label{eq:Uc}
  \begin{array}{lcl}
    (U^c)_{\l\l;\l\l}=-U, &\quad & (U^c)_{\l\l';\l\l'}=-2V, \\
    (U^c)_{\l\l;\l'\l'}=V-\frac{3}{2}J, &\quad & (U^c)_{\l\l';\l'\l}=-2J',
  \end{array} \\ 
  & & 
  \label{eq:Us}
  \begin{array}{lcl}
    (U^s)_{\l\l;\l\l}=U, &\quad & (U^s)_{\l\l';\l\l'}=J, \\ 
    (U^s)_{\l\l;\l'\l'}=\frac{1}{2}J-V, &\quad & (U^s)_{\l\l';\l'\l}=2J',
  \end{array}
\end{eqnarray}
where $\l\ne\l'$.  When the largest eigenvalue of
$\underline{U}^{(c,s)}\underline{\chi}_0(\bq,\Omega=0)$ is
$\lambda^{(c,s)}(\bq)=1$, the response function diverges, indicating
an instability towards an ordered state, with
corresponding eigenvectors, $f^{(c,s)}_{\l\l'}(\bq)$.  Figure
\ref{fig:rpa} shows the largest eigenvalues, $\lambda^c(\bq)$ and
$\lambda^s(\bq)$ along high symmetry lines in the Brillouin zone (BZ)
for ${\bar{n}=0.3}$, $U/t=2$ and temperature $T/t=0.02$, demonstrating the emergence of
various instabilities as we vary Hund's coupling.
For our choice of experimentally motivated parameters, the leading
instabilities are nearly orbital diagonal, ${f^{(c,s)}_{\l\l'}(\bq)\approx
f^{(c,s)}_{\l}(\bq)\delta_{\l\l'}}$.

When $J=0$, the Hubbard interaction $U$ drives
a leading instability in the spin
channel, with $\bq$ generically incommensurate (close to the $M$ points of the BZ
for the range of densities investigated), with all the weight on a single orbital. 
This instability indicates a
tendency towards incommensurate spiral or commensurate stripe antiferromagnetic 
(AF) order -- for each orbital in a different direction.
In the charge channel, we find a subleading
instability, with two degenerate eigenvalues at $\bq=0$, indicating a tendency 
towards ferro-orbital order which will lead to nematicity associated with
broken lattice rotational symmetry. With increasing $J$, the AF
instability gives way to 
a ferromagnetic (FM) instability seen in the spin response at the $\Gamma$-point.
At the same time, the ferro-orbital response is strongly suppressed. Below, we use
mean field theory (MFT) in order to further characterize the broken symmetry phases.

{\it Mean field theory. ---} We study the phase
diagram of our model, Eqns.~(\ref{eq:H0})-(\ref{eq:Hint}), using
a momentum space MFT within a single-$\bQ$ spiral ansatz with a
spatially uniform but orbital-dependent density. This is
captured by a mean field Hamiltonian, ${H_{\rm var}= 
H_0 - \sum_{i\l} (\phi_{\ell}+\mu) n_{i\l}- \sum_{i\ell}{\bf b}_{\ell} \cdot {\bf S}_{i \l}
{\rm e}^{\i\bQ\cdot \br_i}}$, which we use
to generate variational ground states $\ket{\psi_{\rm var}}$ at the desired
charge density by tuning $\mu$. The fields $\phi_{\l},{\bf{b}}_{\l}$, 
and the wavevector $\bQ$ are selected to minimize $\bra{\psi_{\rm var}} H_0+H_{\rm int} \ket{\psi_{\rm var}}$
(see SM).

Fig.~\ref{fig:phasediagram} (a)-(c) shows the MFT phase diagrams for
densities $\bar{n}=0.15,0.30,0.45$. Broadly, we find three phases consistent with RPA: (i) a
paramagnetic (PM) metal where $\phi_\ell\!=\!0, {\bf b}_{\ell}\!=\!0$, (ii)
a commensurate stripe-AF (AF-1) or incommensurate spiral (ICS) metal driven by $U$, where
$\phi_\ell \neq 0$ and ${\bf b}_\ell \neq 0$, which has higher density in
one of the three orbitals,
and (iii) a FM metal driven by Hund's coupling where
${\bf b}_{\ell}\!=\!{\bf b}$, and $\bQ\!=\!0$, either having the same density in all
orbitals (FM), or with one orbital having a lower density than the other two
(i.e., a nematic FM: ${\cal N}$FM). Interestingly, we find a direct (generically
first-order) transition 
from the PM into the AF-1 phase for all three densities, in contrast to the 
RPA which
finds AF-1 only at a fine-tuned density.

To go beyond the single-$\bQ$ ansatz, we 
have also studied the commensurate AF by minimizing the free energy
at $T\!=\! 0.02 t$ assuming a $2\!\times\! 2$ unit cell.
The result for $\bar{n}\!=\! 0.3$
is shown in Fig.~\ref{fig:phasediagram}(d) (see also SM); we find reasonably 
good agreement with the
single-$\bQ$ MFT, but discover small regimes where the single-$\bQ$
order gives way to multi-$\bQ$ condensates where two (AF-$2$) or
three (AF-$3$) wavevectors are simultaneously present. 
A robust feature 
is the presence of simultaneous AF and ferro-orbital order in the AF-$1$ and ${\text{AF-}2}$ phases.
A similar competition between single-$\bQ$ and multi-$\bQ$ phases also appears in single-orbital
honeycomb and triangular lattice Hubbard models
\cite{chern_broken_2012,nandkishore_itinerant_2012}.

The ${\cal N}$FM, AF-$1$, AF-$2$, and ICS phases feature discrete ferro-orbital order which breaks the
hexagonal lattice symmetry. Thus, even if fluctuations melt the magnetic order itself, there may be large 
regimes in the phase diagram where the electronic nematicity survives. Below we use
Landau theory to further understand this interplay of magnetism and nematicity.



{\it Effective field theory. ---} Landau theory is a
powerful tool to analyze magnetic orders
\cite{banerjee2013ferromagnetic} and study spin textures such as
skyrmions which might arise at the (001) LAO-STO interface
\cite{XLi2014}. For the (111) 2DEG, our RPA analysis suggests that the soft
electronic modes include a complex nematic charge mode
${\psi_n = \delta\rho_{xy}+\omega \delta\rho_{yz}+\omega^2
  \delta\rho_{zx}}$ (with $\omega={\rm e}^{\i 2\pi/3}$) constructed
from the slowly varying orbital densities $\delta\rho_\ell$ at the
$\Gamma$-point, and the spin modes, $\vec \varphi_0$ at the
$\Gamma$-point, and $\vec \varphi_\alpha$ at the magnetic wavevectors
$\bQ_\alpha$ ($\alpha=1,2,3$), which can describe both FM and AF
orders.  Since the interorbital hopping is small, $\vec
\varphi_{1,2,3} \sim \vec \varphi_{xy,yz,zx}$ but with weak orbital
admixture.  The spin modes $\vec \varphi_\alpha$ are complex for
incommensurate $\bQ_\alpha$, and real if $\bQ_\alpha$ correspond to
the commensurate $M$-points.
The nematic order parameter $\psi_n$ transforms under anticlockwise
lattice ${\pi/3\text{-rotations}}$ as $\psi_n \to \omega^2 \psi_n$, and under
reflections about the ${\hat{x}\text{-axis}}$ as $\psi^\pdg_n \to
\psi^*_n$. Turning to the spin modes, $\vec \varphi_0$ is invariant
under lattice symmetries, anticlockwise $\pi/3$-rotations leads to
${\vec \varphi_{1} \to \vec\varphi_{2}, \vec \varphi_{2} \to
\vec\varphi_{3}, \vec \varphi_{3} \to \vec\varphi^*_{1}}$.  Under spin
rotations, $\psi_n$ is invariant but all spin modes undergo $SO(3)$
rotations, $\vec\varphi_{0,\alpha} \to \mathfrak{R}
\vec\varphi_{0,\alpha}$. Time-reversal sends $\vec\varphi_{0,\alpha}
\to - \vec\varphi_{0,\alpha}$. Armed with this, the mean field Landau
free energy is ${\cal F} = \int \mathrm{d}^2 x\;( {\cal L}_\psi + {\cal
  L}_\varphi + {\cal L}_{\psi\varphi})$, with 
\bea
\!\!{\cal L}_\psi \!&=&\! r_\psi |\psi_n|^2 + w_\psi ({\psi^3_n} + {\psi^{*3}_n}) + u_\psi |\psi_n|^4 + \ldots\\
\!\!{\cal L}_\varphi \!&=&\! r_0 \vec\varphi_0 \! \cdot \!
\vec\varphi_0 \!+\! r_Q \sum_\alpha |\vec\varphi_\alpha|^2 \!+\!
\ldots \\ 
\!\!{\cal L}_{\psi\varphi} \!&=&\! - \lambda_1 (\psi_n^* {\cal S}^\pdg_n +
\psi_n^\pdg {\cal S}_n^* ) - \lambda_2 |\vec \varphi_0|^2 |\psi_n|^2,
\eea 
where we have defined a complex
magnetic nematic order ${\cal S}_n=|\vec\varphi_1|^2 + \omega
|\vec\varphi_2|^2 + \omega^2 |\vec\varphi_3|^2$ which transforms
analogous to $\psi_n$.  In this effective field theory, the
PM phase corresponds to $(r_{\psi},r_0, r_Q >0)$, the FM
phase corresponds to $(r_0<0, r_\psi, r_Q >0)$, and the various AF
phases correspond to $(r_Q<0,r_0>0)$. Note that we never find a ground
state nematic charge order unaccompanied by spin order in the MFT. The
various types of AF orders will be dictated by higher order (quartic
and sixth order) terms denoted above by ellipsis. In turn, this can
lead to a `pinning field' for the charge nematic order via the cubic
interaction $\lambda_1$ in ${\cal L}_{\psi\varphi}$; our mean field results
indicate $w_\psi, \lambda_1, \lambda_2 > 0$.  Below, we discuss some implications of this
Landau theory, deferring its microscopic derivation to a future publication.

{\bf (a) Incommensurate spiral (ICS):} For the generic incommensurate
instability, the simplest
spin order (which will not lead to any charge modulation) is a
single-mode coplanar spiral at one $\bQ_\alpha$, with the complex
${\vec\varphi_\alpha = \hat{\Omega}_1 + \i \hat{\Omega}_2}$ and
$\hat{\Omega}_1 \cdot \hat{\Omega}_2 =0$. This leads to ${{\cal S}_n
  \sim \omega^{\alpha-1}}$, so the cubic interaction $\lambda_1>0$ will
pin $\psi_n \sim \omega^{\alpha-1}$, causing density enhancement
in orbital $\alpha$.

{\bf (b) Commensurate AF:} For commensurate stripe AF,
$\bQ_\alpha \equiv M_\alpha$, in which case $\vec\varphi_\alpha$ are
real fields.
This case, for which we have also carried out a real space MFT, leads to
three orders.  (i) AF-1 has condensation at a single $M_\alpha$, which
corresponds to a collinear stripe order with $\vec\varphi_\alpha \sim
\hat{\Omega}_1$. This has ${\cal S}_n \sim \omega^{\alpha-1}$, which
pins ${\psi_n \sim \omega^{\alpha-1}}$, leading to a charge nematic
order similar to the ICS state.  (ii) AF-2 features condensation at a
pair of wavevectors $M_\alpha,M_{\beta}$, with
$\la\vec\varphi_\alpha\ra\sim \hat{\Omega}_1$ and
$\la\vec\varphi_{\beta}\ra\sim \hat{\Omega}_2$. This state can be either
collinear, $\hat\Omega_1=\hat\Omega_2$, or coplanar, $\hat{\Omega}_1
\cdot\hat{\Omega}_2 =0$. In both case, however, ${{\cal S}_n, \psi_n
\sim \omega^{\alpha-1} + \omega^{\beta-1}}$, displaying charge nematic
order. (iii) Finally, AF-3 is a triple-$\bQ$ spin crystal, similarly
featuring either collinear or non-coplanar tetrahedral order of the
spins. Both cases are obtained by condensation at all three
$M$-points, with $\la\vec\varphi_{1,2,3} \ra\sim
\hat{\Omega}_{1,2,3}$, and ${\cal S}_n=0$, so no charge nematic is
induced.  In the collinear case $\hat\Omega_1=\hat\Omega_2=\hat
\Omega_3$, while in the non-coplanar spin order $\hat{\Omega}_1 \cdot
\hat{\Omega}_2 =0$, and $\hat\Omega_3 =\pm \hat{\Omega}_1 \times
\hat{\Omega}_2$.  The latter case, in the presence of interorbital
hopping, will feature an anomalous Hall effect
\cite{Martin2008,Kato2010}. The collinear AF-2 and AF-3
will also break translational symmetries with associated charge
modulation driven by terms proportional to $\vec \varphi_\alpha \cdot
\vec \varphi_{\beta} \neq 0$; such orders may be favored by
repulsive interactions between neighboring sites.

{\bf (c) Nematic FM (${\cal N}$FM):} Starting with a uniform FM state, the quartic coupling $\lambda_2>0$,
if sufficiently large,
can drive charge nematicity, since it can change the `mass' of the nematic
field to $(r_\psi - \lambda_2 |\vec \varphi_0|^2)$. This coupling between
FM and nematic orders is not linear in $\psi_n$, unlike the above AFM/ICS cases. Thus, 
the nematicity in this case is not symmetry-enforced. For $w_\psi > 0$, the ${\cal N}$FM will
have depletion of the density in one orbital, as we find from the MFT. 
 
{\it Fluctuation/disorder effects. ---} In 2D, without SOC, thermal
fluctuations will destroy long-range magnetic order at any $T >0$. In
this case, ICS, the ${\cal N}$FM, the collinear ${\text{AF-}1}$, and orthogonal AF-2, will melt
into a charge nematic, reflected in a nonzero $\psi_n, {\cal S}_n$,
which will undergo symmetry restoration via a $Z_3$ clock (or
$3$-state Potts) transition. Within MFT this is a first order
transition, but thermal fluctuations render it a continuous transition
\cite{wu_potts_1982}.  
The non-coplanar AF-3 state will lead to a
magnetically melted state with only chiral order (linked to the $\pm$
choice of $\hat{\Omega}_3$), featuring a nonzero anomalous Hall
effect that vanishes above an Ising transition at which
time-reversal symmetry is restored \cite{Kato2010}.

We speculate that disorder might also weakly suppress long-range magnetic
order, even with SOC, leaving vestigial nematic order \cite{Nie2014}
down to $T=0$. This suggestion is motivated by
Sr$_3$Ru$_2$O$_7$, where the observation of nematic transport
\cite{Borzi2014} near the metamagnetic critical point was recently attributed,
via neutron scattering, to arise from {\it nearly} ordered
SDW phases \cite{Hayden2015}.

{\it Discussion. ---} The electronic nematic phases we have proposed
in (111) 2DEGs will lead to transport anisotropies.  On
symmetry grounds, the scaled resistive anisotropy
$(\rho_{xx}\! -\! \rho_{yy})/(\rho_{xx}\!+\! \rho_{yy})$ will track the nematic
order parameter \cite{Fradkin2000,Carlson2017}. A simple Drude
picture (see SM) shows that, for the coordinates used above,
$\rho_{xx}\!-\! \rho_{yy}\!\sim\! {\rm Re}\,\psi$ while
$\rho_{xy}\!=\! \rho_{yx}\!\sim\! {\rm Im}\,\psi$.
Further signatures of nematic order may be observed in Friedel
oscillations which can be probed using scanning tunneling spectroscopy.  
Even in a conventional phonon-induced SC state, the presence of such
background nematic order would lead to an
anisotropy of the vortex shape and the mobility as well as the critical
current, explaining the anisotropy observed in the resistive transition into the SC state
\cite{davis_superconductivity_2017}.  If such orbital order is weak,
it will be less evident in ARPES
\cite{rodel_orientational_2014, mckeown_walker_control_2014} than
in transport probes. 

Rashba SOC \cite{held_001,Park_PRB2013,Park_PRB2012,Park_PRL2011}
does not
significantly impact the (111) FS for relevant densities $\bar{n} \sim 0.3$, 
or lead to a significant spin-splitting near the
tips of the elliptical FSs where
we find the magnetic instability (see SM)
since orbital mixing is negligible at those momenta. Thus, we expect SOC will not
significantly modify the phase diagram at these densities; however, it can pin the magnetic order 
or convert the uniform FM into a long wavelength
spiral \cite{banerjee2013ferromagnetic,XLi2014}. SOC will have a more significant
impact on low density 2DEGs, and transport properties which
average over the entire FS of all bands.


Our work has not taken into account random
oxygen vacancies - these can locally pin the nematic order but cannot
induce anisotropies in macroscopically averaged 
resistivity measurements. However, such `nematogen'
defects could amplify weak resistive anisotropies, both of the bulk tetragonal phase
in dilute 2DEGs,
as well as of the higher density nematic phases with orbital order.
This interplay, which has been studied in the pnictides \cite{Gastiasoro_PRL2014},
would be worth exploring in the oxide 2DEGs.

Finally, tetragonal lattice distortions, described by a
3-state Potts theory \cite{aharony_trigonal--tetragonal_1977} similar
to ${\cal L}_\psi$, will couple linearly to the nematic
order parameter $\psi_n$ itself, affecting the SDW degrees of freedom as
well. For instance, a tetragonal distortion with elongation of the ${c\text{-axis}}$
will favor the nematic order associated with the
AF-1 state. The interplay of electronic
nematicity explored here, with anisotropies induced by surface
phonon mechanisms, is an important topic for future research.

We thank F. Y. Bruno, Q. Li, L. Miao, A. Chubukov, and E. Fradkin for useful
discussions and feedback. This research was supported by NSERC of Canada.
AP acknowledges the support and hospitality of the International Center for
Theoretical Sciences (Bangalore) during completion of this manuscript.


%

\begin{widetext}
	\section{Supplemental Material}

\subsection{Details of RPA calculation}

The bare susceptibility, which is the building block of the RPA
approximations, can be written in terms of the single particle Green's
functions as
\begin{equation}
\label{eq:chi0G}
\left(\chi_0(\bq,\Omega)\right)_{\l_1\l_2;\l_3\l_4} 
= \frac{1}{N\beta}\sum_{\bk,\omega_n}
G_{\l_2\l_3}(\bk,\omega_n)G_{\l_4\l_1}(\bk+\bq,\omega_n+\Omega)
\end{equation}
Given the unitary matrix $\underline{u}(\bk)$ which diagonalizes
$\underline{h}(\bk)$, and its eigenvalues $\varepsilon_m(\bk)$, the
Green function is
\begin{equation}
\label{eq:G0}
G_{\l\l'}(\bk,\omega_n)=\sum_{m=1}^3\frac{u_{\l m}(\bk)u_{\l' m}^*(\bk)}
{\i\omega_n-\varepsilon_m(\bk)+\mu}.
\end{equation}
Using this we compute the RPA response ${\underline{\chi}^{(c,s)}_{RPA}(\bq,\Omega)=\underline{\chi}_0(\bq,\Omega)
	(1-\underline{U}^{(c,s)}\underline{\chi}_0(\bq,\Omega))^{-1}}$. In order to understand the instabilities, it suffices to focus on the
matrix product $\underline{U}^{(c,s)}\underline{\chi}_0(\bq,\Omega=0)$; if its eigenvalues exceed unity, it signals an RPA instability.
The RPA results for different electron densities are presented in figure \ref{fig:rpa2}. In the small $J$ regime and for small densities, the leading instability is a $2k_F$ instability in the spin channel in the $\Gamma\to$ M direction, which is commensurate for specific densities and incommensurate in general. The $\bar{n}=0.3$ electrons per site is an example of a commensurate order, which we study in the main text using a $2\times2$ unit cell. For larger densities, when $2k_F$ exceeds the length of the reciprocal lattice vectors, the ordering wavevector moves to the M $\to$ K direction. When Hund's coupling is large, we get a FM transition corresponding to a leading instability at the $\Gamma$ point.
\begin{figure}[h]
	
	\begin{subfigure}[b]{0.32\textwidth}
		\centering
		\includegraphics[width=\textwidth]{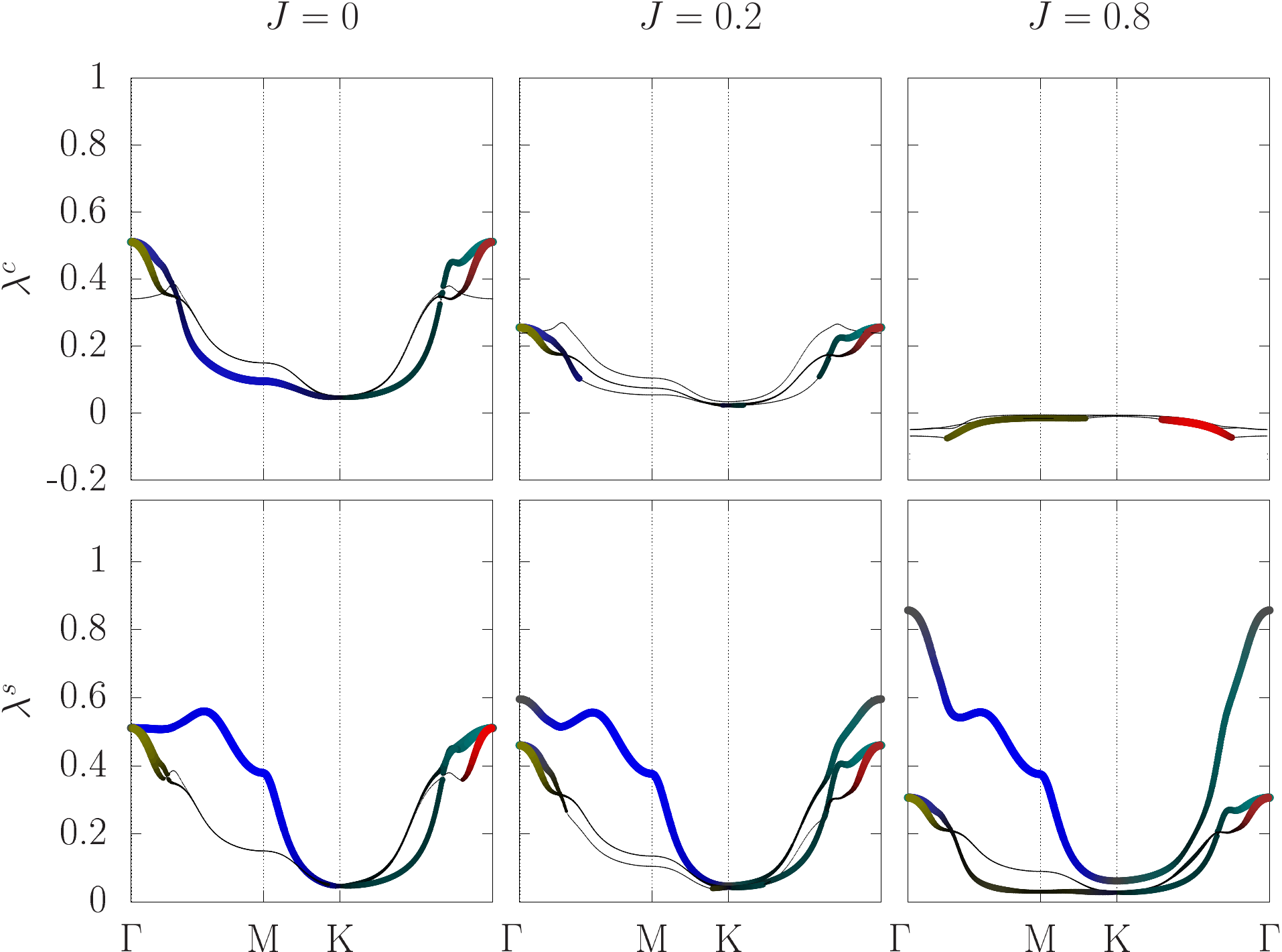}
		\caption{$\bar{n}=0.15$}
	\end{subfigure}
	\begin{subfigure}[b]{0.32\textwidth}
		\centering
		\includegraphics[width=\textwidth]{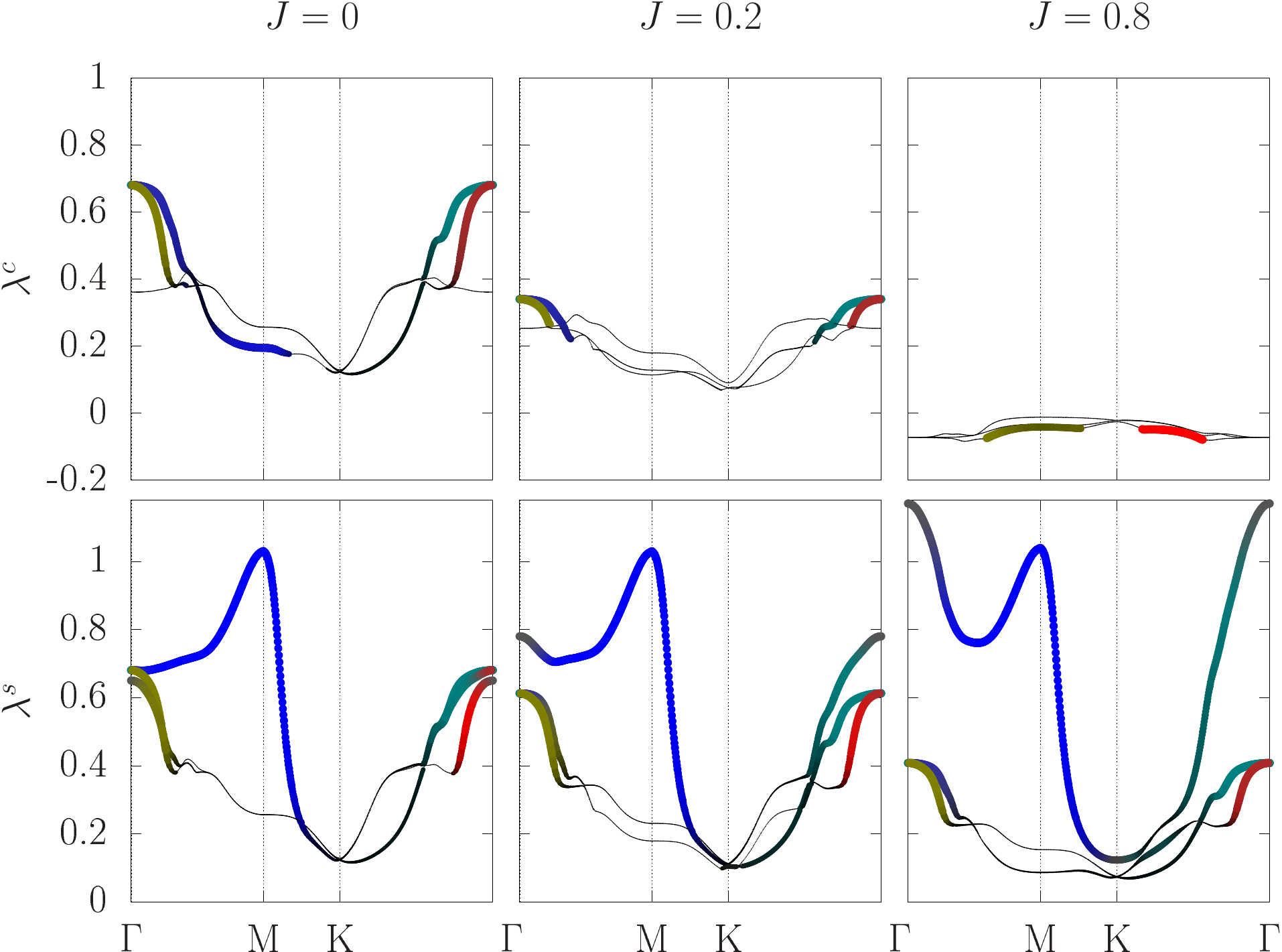}
		\caption{$\bar{n}=0.30$}
	\end{subfigure}
	\begin{subfigure}[b]{0.32\textwidth}
		\centering
		\includegraphics[width=\textwidth]{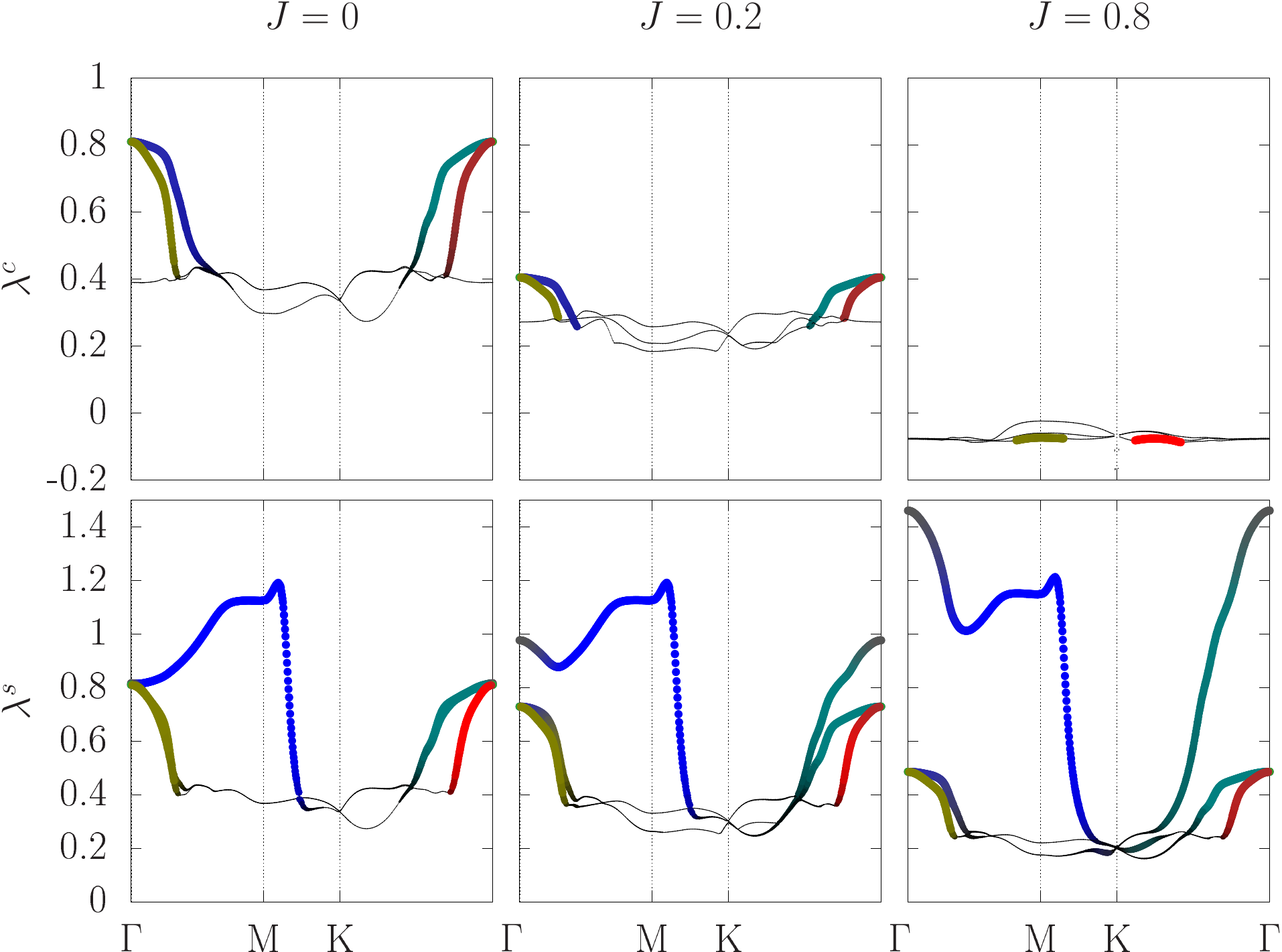}
		\caption{$\bar{n}=0.45$}
	\end{subfigure}
	\caption{Dominant three eigenvalues $\lambda^{(c,s)}$ ($c$=charge, $s$=spin) of the
		matrix product
		$\underline{U}^{(c,s)}\underline{\chi}_0(\bq,\Omega=0)$ for densities (a) $\bar{n}=0.15$, (b) $\bar{n}=0.30$, and (c) $\bar{n}=0.45$
		and indicated Hund's coupling.
		The color indicates the orbital content: red=$xy$, green=$xz$, blue=$yz$.}
	\label{fig:rpa2}
\end{figure}	
\subsection{Spin-orbit coupling}
Surface and interface 2DEGs are subject to Rashba spin-orbit coupling (SOC). This arises from a combination of inversion symmetry breaking at the interface, combined with atomic SOC.\cite{Park_PRL2011,Park_PRB2012,Park_PRB2013}
The inversion breaking term is odd under $\bk\rightarrow-\bk$, and can be phenomenologically incorporated by allowing for inter-orbital hoppings which are ordinarily forbidden by the presence of inversion symmetry.  
Such terms have been labelled `orbital Rashba' terms.\cite{Park_PRL2011,Park_PRB2012,Park_PRB2013} For example, an electron in the $xy$ orbital could not move to a $yz$ orbital along the $\hat{a}$ direction if inversion symmetry is not broken. However, the broken symmetry at the interface will distort the orbital cloud, leading to an inter-orbital overlap which will be either stronger or weaker depending on the hopping direction (e.g., an 
$xy\rightarrow yz$ hopping is strong in the $-\hat{a}$ direction but weak in the $+\hat{a}$ direction), giving rise to inter-orbital hopping terms $\propto \sin(\bk\cdot\textbf{r})$. When combined with atomic SOC, this leads to the Rashba effect which has been shown to play a role in (001) oxide 2DEGs. Here, we generalize this idea to the (111) 2DEG.

Explicitly, for the (111) 2DEG in momentum space, we get for the inversion breaking `orbital Rashba' term $H_{OR} = \sum_{\bk,\ell\ell',\sigma}c^\dagger_{\ell\sigma}(\bk)h^R_{\ell\ell'}(\bk)c_{\ell'\sigma}(\bk)$, where
\begin{equation}
h^{R}(\bk) = 2 \i \lambda_{OR} \begin{pmatrix}
0 & \sin(k_b)-\sin(k_c) & \sin(k_a)+\sin(k_c) \\
\sin(k_c)-\sin(k_b) & 0 & \sin(k_b)-\sin(k_a) \\
-\sin(k_a)-\sin(k_c) & \sin(k_a)-\sin(k_b) & 0 \\
\end{pmatrix}
\end{equation}
and $\ell \equiv (yz,zx,xy)$.
Atomic spin-orbit coupling takes the form $H_{A}= - \lambda_A\;\vec{L}\cdot\vec{S}$ with $\lambda_A\approx20$ meV (see e.g., Ref.\onlinecite{held_001}). In the $(c_{yz,\upa},c_{zx,\upa},...,c_{xy,\downarrow})^T$ basis, we can write this as $H_{A} =  \i \frac{\lambda_A}{2} \sum_{\bk} \varepsilon^\pdg_{\ell m n} \tau^n_{\sigma\sigma'} 
c^\dg_{\ell \sigma}(\bk) c^\pdg_{m \sigma'}(\bk)$, where $\varepsilon_{\ell m n}$ is the Levi-Civita symbol and $\tau^n$ the Pauli matrices.

\begin{figure}[h]
	\begin{subfigure}[b]{0.3\textwidth}
		\centering
		\includegraphics[width=\textwidth]{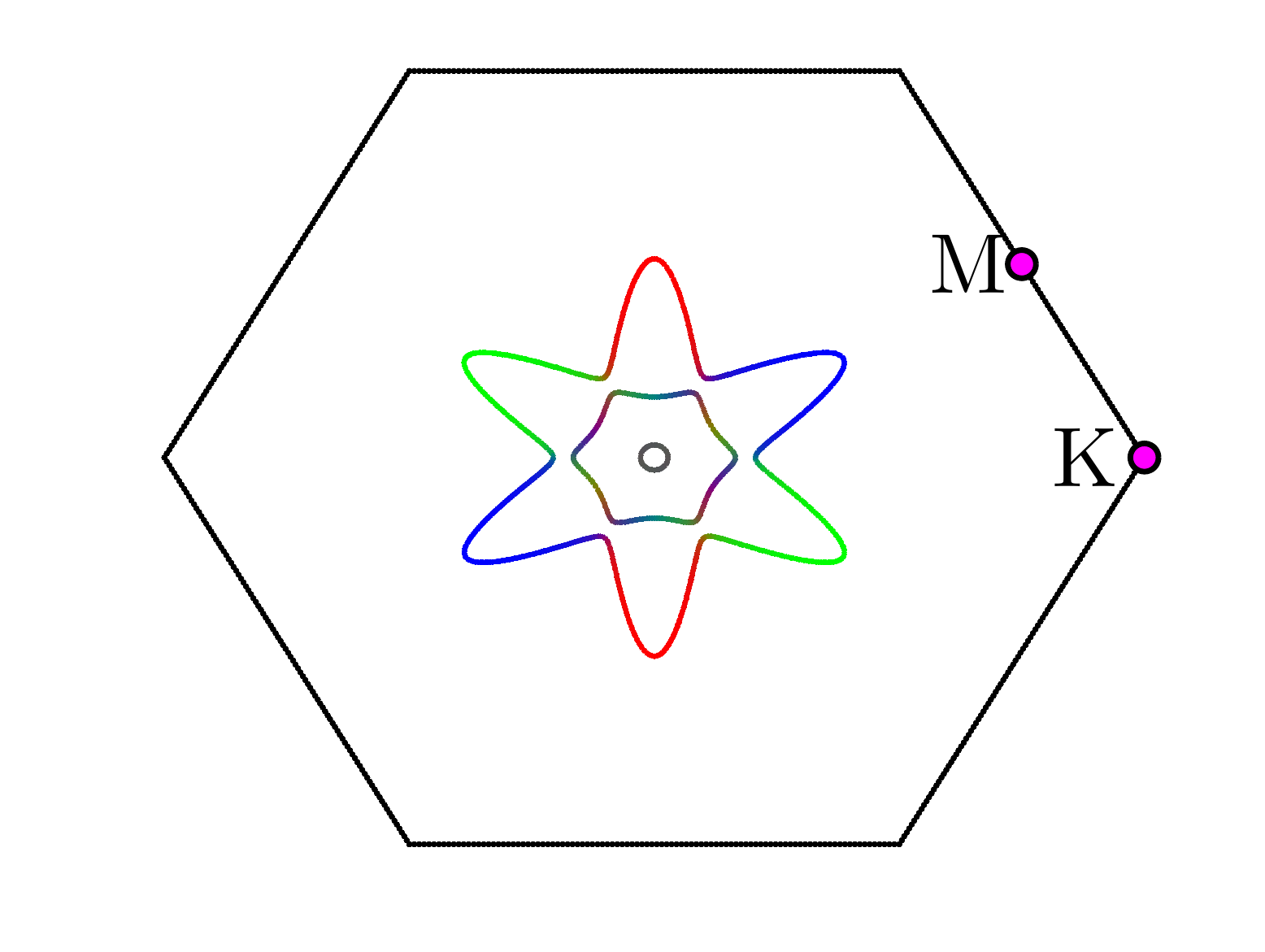}
		\caption{$\lambda_{OR}=0, \lambda_A=0$. }
		\label{fig:0}
	\end{subfigure}
	\begin{subfigure}[b]{0.3\textwidth}
		\centering
		\includegraphics[width=\textwidth]{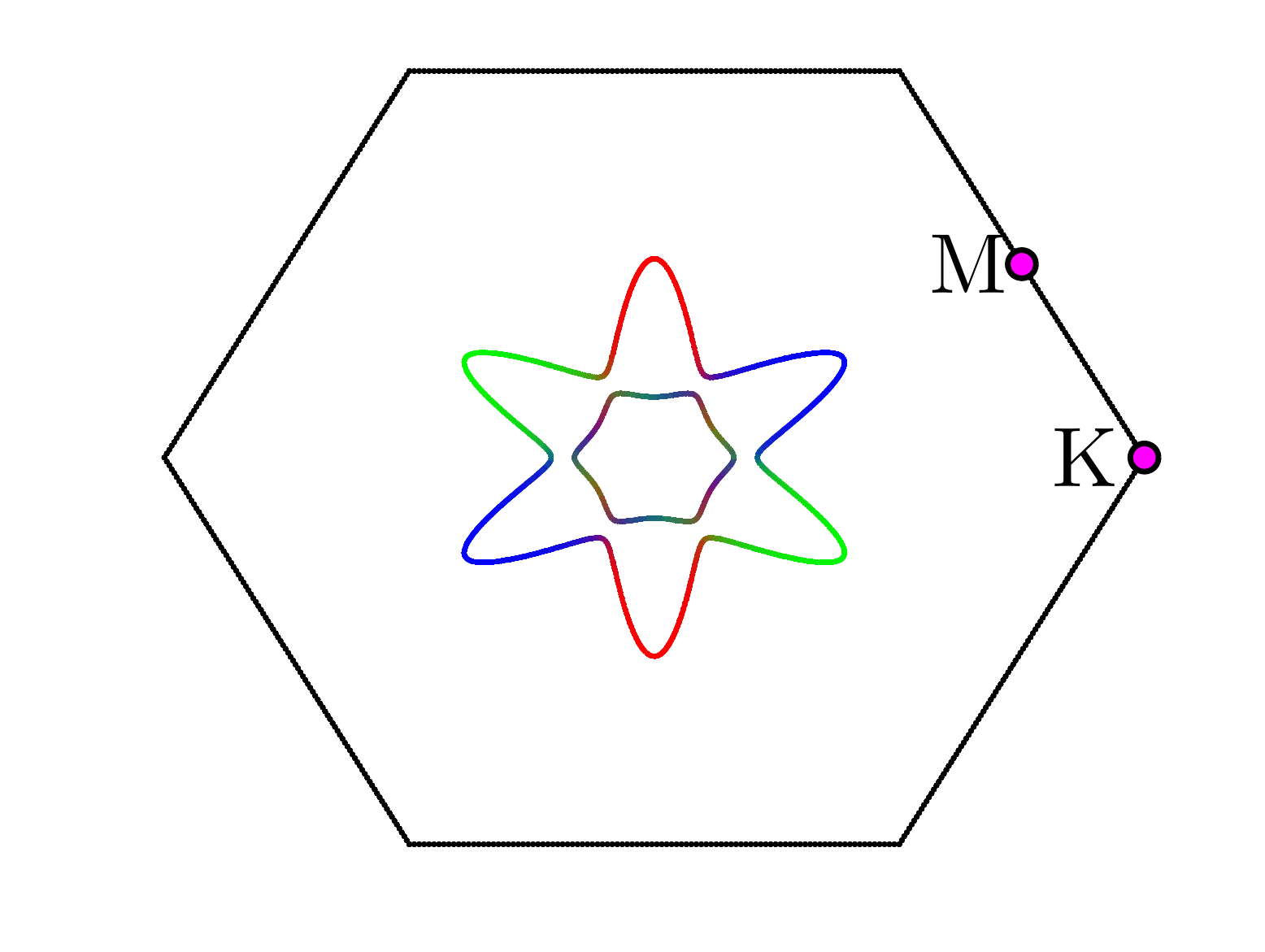}
		\caption{$\lambda_{OR}=0, \lambda_A=20$ meV. }
		\label{fig:A}
	\end{subfigure}
	\begin{subfigure}[b]{0.3\textwidth}
		\centering
		\includegraphics[width=\textwidth]{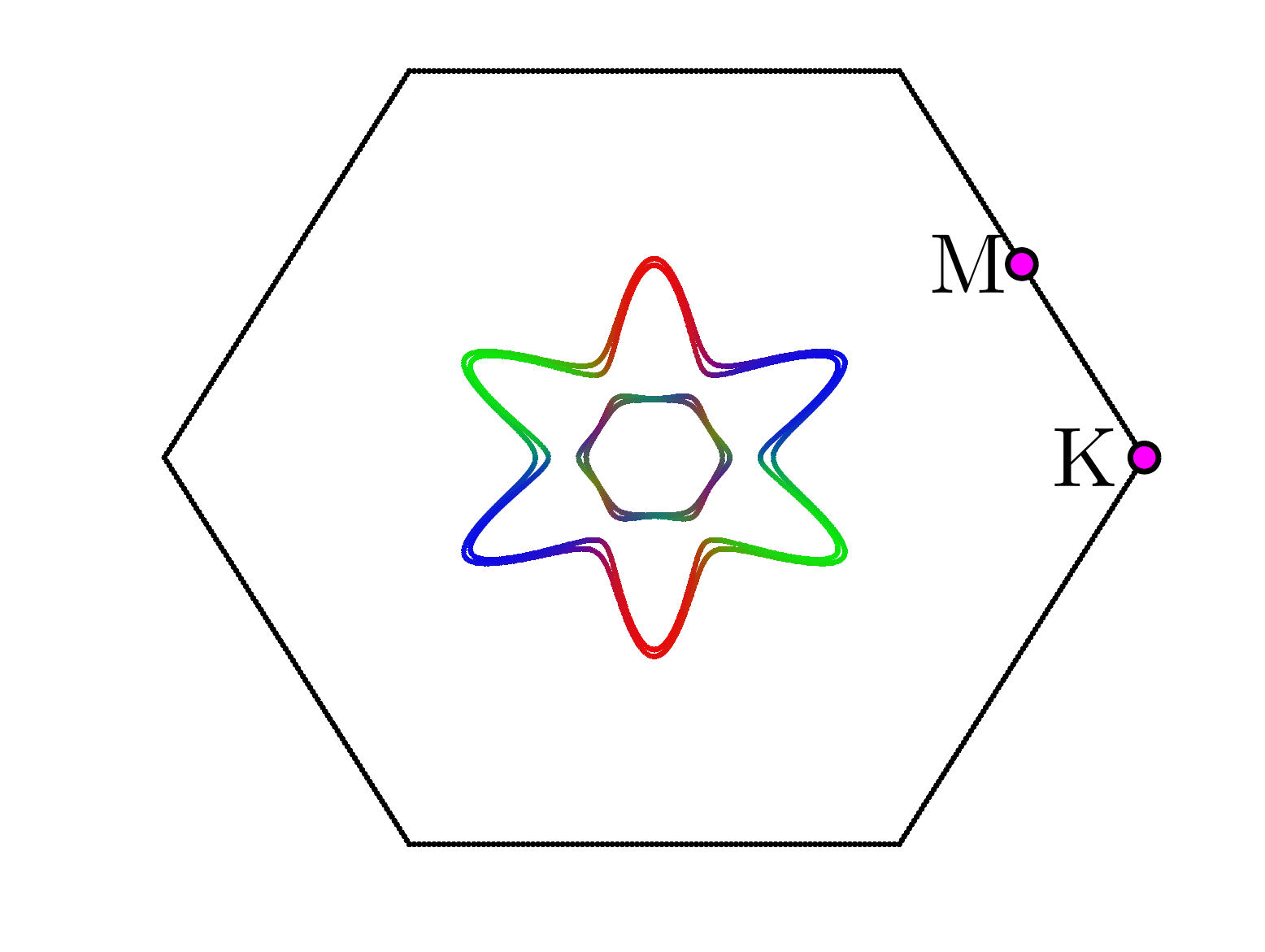}
		\caption{$\lambda_{OR} = \lambda_A=20$ meV. }
		\label{fig:AR}
	\end{subfigure}
	\begin{subfigure}[b]{0.3\textwidth}
		\centering
		\includegraphics[width=\textwidth]{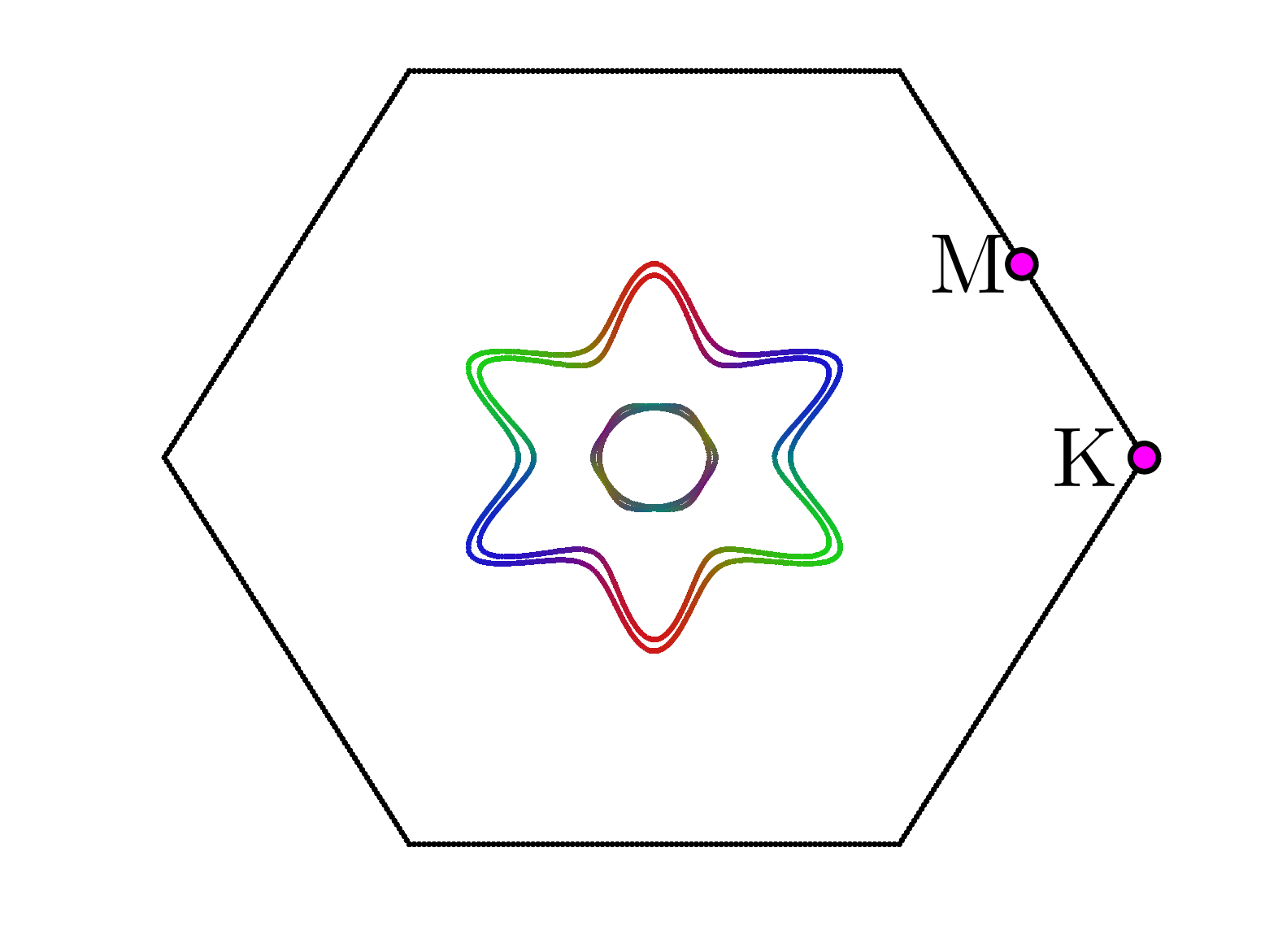}
		\caption{$\lambda_{OR}=40$ meV, $\lambda_A=20$ meV. }
		\label{fig:0612}
	\end{subfigure}
	\begin{subfigure}[b]{0.3\textwidth}
		\centering
		\includegraphics[width=\textwidth]{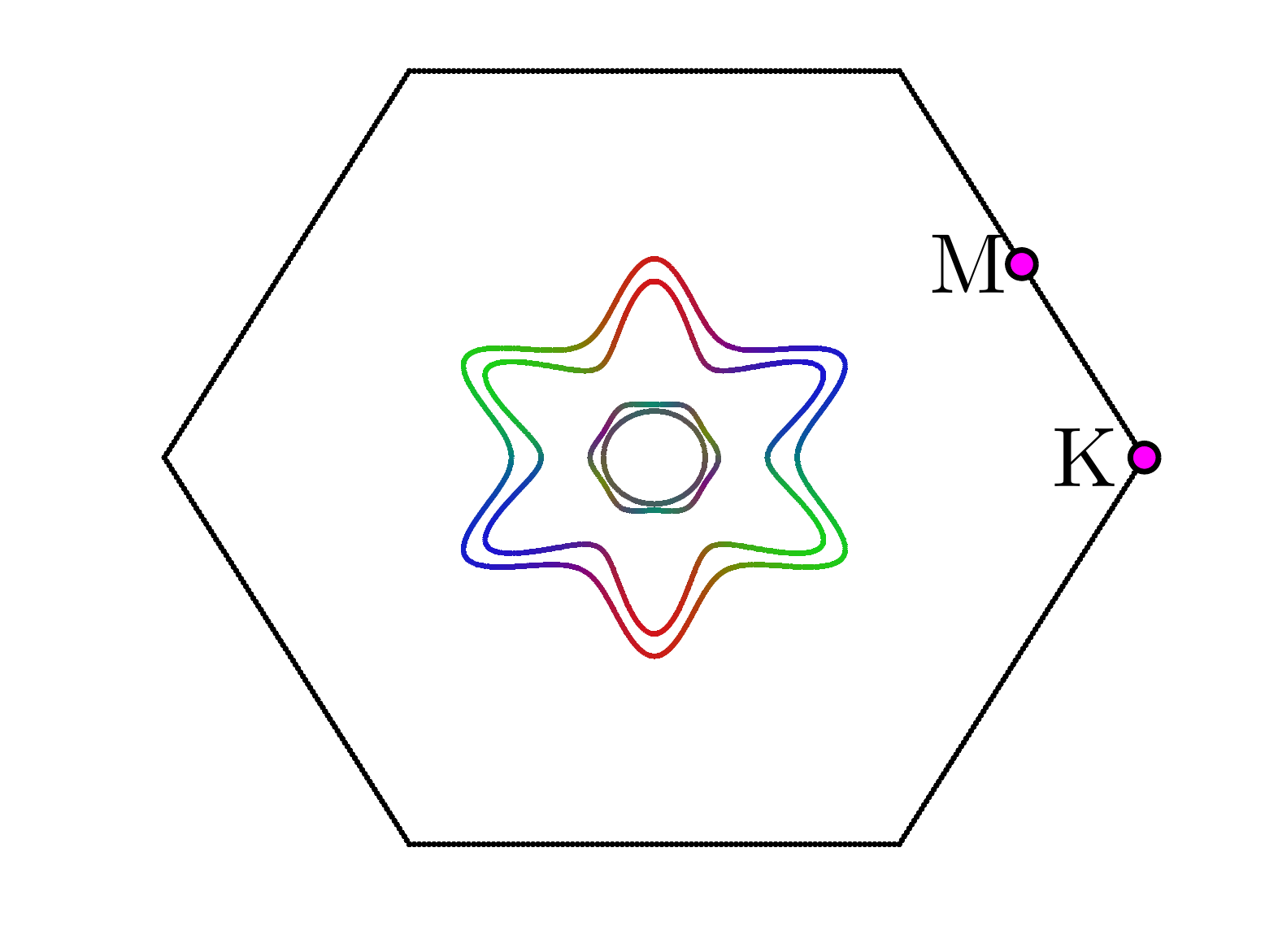}
		\caption{$\lambda_{OR}=\lambda_A=40$ meV. }
		\label{fig:012012}
	\end{subfigure}
	\caption{Fermi surfaces for different strengths of orbital Rashba $\lambda_{OR}$ coupling and atomic SOC $\lambda_A$ at $\bar{n}=0.3$.}
\end{figure}

\noindent What is the impact of such atomic SOC and inversion breaking on the Fermi surface? Comparing figure \ref{fig:0} ($\lambda_A=0, \lambda_{OR}=0$) to figure \ref{fig:A} ($\lambda_A=20$ meV, $\lambda_{OR}=0$) we see that the main effect of atomic spin-orbit coupling on the Fermi surface is the disappearance of the small
band near the $\Gamma$ point, which gets pushed below the Fermi level. The leading magnetic instability uncovered in our RPA (reported in the main
text and above) occurs at the wavevector connecting the tips of the elliptical Fermi surfaces. Near these tips, we find that the Fermi surface is nearly unaffected
which we can understand since the Fermi surface near these regions is almost entirely composed of a single orbital.
Using $\lambda_A=\lambda_{OR} = 20$ meV (see Ref.\onlinecite{held_001}), we obtain the Fermi surface shown in figure \ref{fig:AR}. We can see that the Rashba spin-orbit coupling breaks the spin degeneracy, previously protected by inversion and time-reversal symmetry. The tips of ellipses are once again nearly unaffected by this term, the main changes to Fermi surfaces being in the $\Gamma\rightarrow\text{K}$ direction where the different orbitals tend to strongly hybridize. However, even here the changes are small.
We include figures \ref{fig:0612} and \ref{fig:012012} to show the effect of even larger inversion breaking 
on the Fermi surfaces; however we point out that these Fermi surfaces 
are not consistent with ARPES data on the STO surface 2DEG, suggesting that they are too large to be of relevance.

\subsection{Details of the MFT calculation}
The mean field theory can be cast in the form of a variational minimization where we pick the variational state to be the ground state of a 
variational Hamiltonian
${H_{\rm var}= 
	H_0 - \sum_{i\l} (\phi_{i \ell}+\mu) n_{i\l}- \sum_{i\ell}{\bf b}_{i \ell} \cdot {\bf S}_{i \l}}$, with $\{\phi_{i \ell}, {\bf b}_{i \ell}\}$ being variational parameters which
allow in general for site and orbital dependent order parameters.
The expression for the variational free energy is given by:
\begin{equation}
\begin{split}
{\cal F}_{\rm var} =& \sum_{i\ell\ell'}\left\{\langle n_{i,\ell}\rangle\langle n_{i,\ell'}\rangle\left[\frac{U}{4}\delta_{\ell\ell'}+\frac{1}{2}\left(U-\frac{5}{2}J\right)(1-\delta_{\ell\ell'})\right]-\langle\mathbf{S}_{i,\ell}\rangle\cdot\langle\mathbf{S}_{i,\ell'}\rangle\left[U\delta_{\ell\ell'}+J(1-\delta_{\ell\ell'})\right]\right\}\\
&+\int_{FBZ}(\mathrm{d}^2\mathbf{k})\mathrm{Tr}(\underline{u}^\dagger(\mathbf{k}) \underline{h}(\mathbf{k})\underline{u}(\mathbf{k})\underline{n_F}(\mathbf{k}))
- T S 
\end{split}
\label{eqn:FE}
\end{equation}
where $\underline{u}(\mathbf{k})$ is the matrix which diagonalizes
$H_{\rm var}$, $\underline{n_F}$ is the Fermi-Dirac distribution, and the thermal expectation values $\la . \ra$ are taken with respect to the
equilibrium density matrix of the variational Hamiltonian, i.e., ${\rm e}^{-H_{\rm var}/T}$, and
$S$ is the fermion entropy for given field configuration $\phi_{i\ell}$ and ${\bf b}_{i\ell}$.

We discuss two cases in the main text, a single-$\bQ$ spiral mean field theory which can be formulated directly 
in momentum space with a single site unit cell, and a case
where we study broken symmetry states with $2\times 2$ real space unit cell which allows us to also explore multi-$\bQ$ order parameters for
the commensurate AF.

\subsubsection{Spiral MFT (zero temperature)}
The $T=0$ spiral mean field theory is performed with only intra-orbital orders, uniform density and spiral magnetization with fixed wavevector $\bq$,
so $\la n_{i\ell\sigma} \ra = \frac{1}{2} \rho_\ell$, and $\la S^{\pm}_{i\ell} \ra = m_\ell {\rm e}^{\pm \i \bq\cdot\br_i}$. This leads to the effective Hamiltonian
\bea
H_{\rm spiral} =  \sum_{\bk} \Psi^\dg(\bk) \mathcal{H}(\bk) \Psi(\bk)
\eea
where $\Psi^\dg(\bk) \equiv (c^\dg_{yz,\upa}(\bk), c^\dg_{zx,\upa}(\bk),c^\dg_{xy,\upa}(\bk),c^\dg_{yz,\dna}(\bk+\bq), c^\dg_{zx,\dna}(\bk+\bq),c^\dg_{xy,\dna}(\bk+\bq))^T$,
and
\bea
\mathcal{H}(\bk) = 
\begin{pmatrix} 
	\varepsilon^{yz}_\bk & \gamma^{yz,zx}_\bk & \gamma^{yz,xy}_\bk & \zeta_{yz} & 0 & 0 \\
	\gamma^{yz,zx *}_\bk & \varepsilon^{zx}_\bk & \gamma^{zx,xy}_\bk &0 & \zeta_{zx} & 0 \\
	\gamma^{yz,xy *}_\bk  & \gamma^{zx,xy *}_\bk & \varepsilon^{xy}_\bk & 0 & 0 & \zeta_{xy} \\
	\zeta_{yz} & 0 & 0 & \varepsilon^{yz}_{\bk+\bq} & \gamma^{yz,zx}_{\bk+\bq} & \gamma^{yz,xy}_{\bk+\bq} \\
	0 & \zeta_{zx} & 0 &  \gamma^{yz,zx *}_{\bk+\bq} & \varepsilon^{zx}_{\bk+\bq} & \gamma^{zx,xy}_{\bk+\bq}\\
	0 & 0 & \zeta_{xy} & \gamma^{yz,xy *}_{\bk+\bq}  & \gamma^{zx,xy *}_{\bk+\bq} & \varepsilon^{xy}_{\bk+\bq} 
\end{pmatrix}
\eea
Here 
\bea
\varepsilon^\ell_\bk &=& (\varepsilon^{\ell (0)}_\bk - \mu) + \frac{U}{2} \rho_\ell + (U- \frac{5}{2} J)   \sum_{\ell \neq \ell'} \rho_{\ell'} \\
\zeta_{\ell}&=& -U m_\ell - J \sum_{\ell \neq \ell'} m_{\ell'}
\eea
with $\varepsilon^{\ell (0)}_\bk$ being the bare intra-orbital dispersion for orbital $\ell$, and $\gamma_\bk$ being inter-orbital hybridization. Using expression \eqref{eqn:FE}, we compute the energy for various $\bq$ and pick the wavevector which minimizes the energy.
This leads us to the phase diagrams shown in Figs.~2(a-c) of the main text.

\subsubsection{MFT with $2\times 2$ real space unit cell}

For the $2\times2$ unit cell, the most general $\bf b$ field can be expressed in terms
of 4 vector parameters ${\mb}_{\alpha}$ ($\alpha=0,1,2,3$),
\begin{equation}
\!\! {\bf b}_{i \l} \!=\! \mb_0\!+\!\mb_1(-1)^{m+n} \! \delta_{\l,yz}\!+\!\mb_2(-1)^{m} \! \delta_{\l,zx}\!+\!\mb_3(-1)^{n} \! \delta_{\l,xy}
\end{equation}
where $(m,n)$ correspond to lattice coordinates of site $i$ along the
$\hat{a}$ and $\hat{c}$ directions respectively. We thus formulate this mean field theory partly in momentum space where we
work in the corresponding reduced BZ.
The paramagnetic state corresponds to
$\mb_\alpha=0$ for all $\alpha$, and the ferromagnetic state to only
$\mb_0 \neq 0$. The different AF states all correspond to $\mb_0=0$
and different nonzero $\mb_\alpha$ (for $\alpha\neq0$).  The collinear
AF-1 (or AF-2 or AF-3) state are obtained when respectively only a
single component of one (or two or three) $\mb_\alpha$ is nonzero. The
orthogonal AF-2 (or AF-3) corresponds to two (three) orthogonal
choices of nonzero $\mb_\alpha$ fields. Minimizing the free energy, we are led to the phase diagram shown in Fig.~2(d) of the main text.

As seen from the figure, there are large regimes where we find robust AF-1 and FM
phases. We also find smaller regimes where AF-2 and AF-3 phases are stabilized which involved the superposition of
two or three $\bQ$ orders, but the free energy per site for these states is lower than the AF-1 state by a very small
amount (of order
$\sim10^{-5}t$). Furthermore, the free energy difference per site between different superpositions (collinear, coplanar, orthogonal)
in the multi-$\bQ$ states (AF-2, AF-3) are even smaller, $\sim10^{-7}t$.

We have studied the temperature dependence of a few points in the AF
phases and found that as the temperature is raised, the AF-2 and AF-3
become an AF-1 phase, consistent with \cite{nandkishore_itinerant_2012}. The mean field critical temperature
into the PM starting significantly deep in the AF and FM phases is
$\simeq0.18t$; without SOC, this will be suppressed to zero, leaving
only vestigial orbital and nematic orders.  A numerical study of the
renormalized $T_c$ for such vestigial orders taking full thermal
fluctuations into account will be discussed elsewhere.

\subsection{Nematicity in the Drude conductivity}

Simple Drude-like considerations are enough to determine how the
electric conductivity depends on the nematic order parameter
$\psi_n$. We begin by assuming that only $\psi_n$ has condensed
and that there is no magnetic order. For simplicity, we further assume
that the inter-orbital hopping is zero. Thus, each orbital, denoted by
	its preferred hopping direction, $\ell=a (xy), b (xz) ,c (yz)$, has a separate
	contribution to the conductivity tensor,
	$\hat\sigma=\sum_\ell\hat\sigma_\ell$. In terms of Cartesian coordinates,
	$x,y$, used in the main text, with $\hat{a}=\hat{x}$, $\hat{b}=\hat{x}/2+\hat{y} \sqrt{3}/2$, and
	$\hat{c}=-\hat{x}/2+\hat{y} \sqrt{3}/2$, we get
\begin{equation}
\label{eq:sigmaabc}
\hat\sigma_a=\left(
\begin{array}{cc}
\sigma_1 & 0 \\ 0 & \sigma_2
\end{array}\right),\quad
\hat\sigma_b 
= \left(
\begin{array}{cc}
\frac{1}{4}\sigma_1+\frac{3}{4}\sigma_2 & 
\frac{\sqrt{3}}{4}(\sigma_1-\sigma_2) \\ 
\frac{\sqrt{3}}{4}(\sigma_1-\sigma_2) & 
\frac{3}{4}\sigma_1+\frac{1}{4}\sigma_2
\end{array}\right),\quad
\hat\sigma_c 
= \left(
\begin{array}{cc}
\frac{1}{4}\sigma_1+\frac{3}{4}\sigma_2 & 
-\frac{\sqrt{3}}{4}(\sigma_1-\sigma_2) \\ 
-\frac{\sqrt{3}}{4}(\sigma_1-\sigma_2) & 
\frac{3}{4}\sigma_1+\frac{1}{4}\sigma_2
\end{array}\right),
\end{equation}
where, in general $\sigma_1\ne\sigma_2$. A six-fold rotation
operation, $C_6$, relates the tensors above:
$\hat\sigma_b=C_6\hat\sigma_aC_6^{-1}$ and
$\hat\sigma_c=C_6^{-1}\hat\sigma_aC_6$. Within a Drude picture, the
matrix elements are proportional to the density of electrons, in this
case, for each orbital separately. Therefore, we can approximate
$\hat\sigma\approx\sum_\ell\rho_\ell\hat\sigma_\ell/\rho_0$, where,
$\rho_\ell$ is the electron density in each orbital, while the average
density per orbital is denoted by $\rho_0$. The nematic order
parameter captures the imbalance in orbital densities,
$\psi_n\sim((\rho_a-\rho_0)+\omega(\rho_b-\rho_0)+\omega^*(\rho_c-\rho_0))/\rho_0$.
Putting everything together, we find
\begin{equation}
\label{eq:sigmapsi}
\hat\sigma\approx  \frac{3}{2}(\sigma_1+\sigma_2 )
\left(\begin{array}{cc}
1 & 0 \\ 0 & 1
\end{array}\right)
+ \frac{1}{2}(\sigma_1-\sigma_2)
\left(\begin{array}{cc}
{\rm Re}\,\psi_n & {\rm Im}\,\psi_n
\\ {\rm Im}\,\psi_n & -{\rm Re}\,\psi_n
\end{array}\right).
\end{equation}
Finally, assuming $\psi_n$ is small, one can obtain a qualitatively
similar dependence of the resistivity tensor $\hat\rho$ on $\psi_n$.
\end{widetext}
\end{document}